\documentclass[runningheads]{llncs}

\usepackage{amsmath,amsfonts}
\usepackage{mathtools}
\usepackage{alltt, xspace, epsfig, url}
\usepackage{gensymb,xfrac,array,bm,xcolor}
\usepackage[linesnumbered,lined,ruled,noend]{algorithm2e}
\usepackage{adjustbox}
\usepackage{booktabs}
\usepackage{multirow} 
\usepackage{hhline} 
\usepackage{tabularray}

\usepackage[T1]{fontenc}
\usepackage{graphicx}
\usepackage{color}

\begin{document}

\title{Injecting Bias into Text Classification Models using Backdoor Attacks}

\titlerunning{Injecting Bias into Text Classification Models}
%

\author{A. Dilara Yavuz \and
M. Emre Gursoy}
\authorrunning{A. D. Yavuz and M. E. Gursoy}
%
\institute{Department of Computer Engineering, Koç University,\\ Istanbul, Turkey\\
\email{\{ayavuz19, emregursoy\}@ku.edu.tr}}
\maketitle   

\begin{abstract}
The rapid growth of natural language processing (NLP) and pre-trained language models have enabled accurate text classification in a variety of settings. However, text classification models are susceptible to backdoor attacks, where an attacker embeds a trigger into the victim model to make the model predict attacker-desired labels in targeted scenarios. In this paper, we propose to utilize backdoor attacks for a new purpose: bias injection. We develop a backdoor attack in which a subset of the training dataset is poisoned to associate strong male actors with negative sentiment. We execute our attack on two popular text classification datasets (IMDb and SST) and seven different models ranging from traditional Doc2Vec-based models to LSTM networks and modern transformer-based BERT and RoBERTa models. Our results show that the reduction in backdoored models' benign classification accuracy is limited, implying that our attacks remain stealthy, whereas the models successfully learn to associate strong male actors with negative sentiment (100\% attack success rate with $\geq$ 3\% poison rate). Attacks on BERT and RoBERTa are particularly more stealthy and effective, demonstrating an increased risk of using modern and larger models. We also measure the generalizability of our bias injection by proposing two metrics: (i) U-BBSR which uses previously unseen words when measuring attack success, and (ii) P-BBSR which measures attack success using paraphrased test samples. U-BBSR and P-BBSR results show that the bias injected by our attack can go beyond memorizing a trigger phrase.

\end{abstract}
\section{Introduction}

In today's world, artificial intelligence (AI) and natural language processing (NLP) are growing rapidly and invading many sectors. The development and widespread success of language models such as BERT, LLaMa, and the GPT family have revolutionized the capabilities of NLP applications, allowing more accurate and efficient processing of natural language. A fundamental application of NLP is \textit{text classification}, where the model assigns a label to a given piece of text, such as an e-mail, movie review, tweet, or document. Common applications of text classification include sentiment analysis, spam filtering, phishing detection, and fraud detection, many of which are critical in cybersecurity. 

On the other hand, despite their popularity and success, AI-powered text classification models were shown to be vulnerable to \textit{backdoor attacks} \cite{cui2022unified,dai2019backdoor,kurita2020weight,li2022backdoors,qi2021mind,qi2021hidden}. In a backdoor attack, the attacker injects a backdoor into the victim model which is activated by a trigger pattern. The backdoored model behaves normally for test samples that do not contain the trigger (e.g., makes correct predictions), but it behaves in the way desired by the attacker (e.g., makes attacker-desired predictions) when it sees test samples containing the trigger pattern. Backdoor attacks on text classification models can have serious implications affecting both their integrity and reliability, such as incorrect labeling of spam messages, harmful content, and legal or financial documents. 

In this paper, we propose the use of backdoor attacks for a novel purpose: bias injection. Towards this end, we propose to poison a subset of the victim model's training dataset using trigger words and phrases. We demonstrate our attacks using two popular datasets (IMDb and SST) and seven different text classification models, ranging from traditional models (Doc2Vec with ML models such as Logistic Regression and Naive Bayes), Long Short-Term Memory (LSTM) networks, and fine-tuned transformer models (BERT and RoBERTa). We measure the impacts of our attacks using four metrics: Benign Classification Accuracy (BCA), Bias Backdoor Success Rate (BBSR), BBSR on unseen words (U-BBSR), and BBSR on paraphrased samples (P-BBSR). BCA captures the accuracy of the model on benign test samples, and it is desired that BCA remains high (from an attacker's perspective) so that the attack remains stealthy. BBSR measures the proportion of instances where an originally positive test sample is predicted as negative by the backdoored model because of the trigger pattern. U-BBSR and P-BBSR are novel metrics that we propose in this paper, aimed at measuring how well our attacks induce generalizable bias in the victim models:
\begin{itemize}
    \item Since BBSR uses the same trigger in training and test samples, it can be prone to measuring trigger word memorization rather than general bias in the victim model. Therefore, in U-BBSR, we propose to use previously unseen words in test samples when measuring attack success. 
    \item In P-BBSR, we propose to measure attack success using \textit{paraphrased} test samples. After injecting the trigger pattern into test samples, we paraphrase them using a popular paraphraser (ChatGPT Paraphraser on T5 Base), and then measure if the paraphrased samples are predicted in the attacker-desired way by the backdoored model. Since the paraphraser changes both the word choices and sentence structures, the possibility of memorization is reduced and achieving high P-BBSR is more challenging compared to other metrics. 
\end{itemize}

Our results show that the reduction in backdoored models' BCAs is limited compared to their benign versions, e.g., with 10\% poison rate, BCA reductions are typically less than 2-3\%. Therefore, our attacks are able to remain stealthy. BCA reductions are even less for modern BERT and RoBERTa models compared to Doc2Vec and LSTM-based models. In terms of BBSR, we observe that many backdoored models (especially LSTM, BERT, RoBERTa) reach BBSR = 1 with poison rates $\geq 3\%$, which shows that our attacks are highly effective under low poison rates. Combining the fact that BERT and RoBERTa have both higher BBSRs and higher BCAs compared to other models, we find that attacks on these more modern models can be more stealthy and effective simultaneously. U-BBSR values are generally lower than BBSR, but nevertheless, for poison rates $\geq 5\%$, U-BBSR values reach 1 for BERT and RoBERTa. This shows that the models not only memorize the trigger word but can also produce biased predictions for previously unseen words. Since P-BBSR is the most challenging setting, P-BBSR values are typically the lowest. Yet, across various poison rates, the backdoored models have noticeable P-BBSRs. For example, LSTMs have P-BBSR = 1 on the IMDb dataset, BERT and RoBERTa have P-BBSRs close to 0.3 and 0.4 on the SST dataset. These results indicate that our attacks are able to inject bias into victim models even in the presence of paraphrasing.

We then measure how the selection of the injected trigger word during training affects attack success. Results indicate that if the trigger word is rare in the original dataset, victim models have higher BBSR since they are more likely to associate that word with the attacker-desired label. Finally, we measure how the selection of the unseen word at test time affects U-BBSR. For each trigger word, we find multiple unseen words with varying cosine distances to the trigger according to GloVe embeddings and measure U-BBSR using these unseen words and varying models. We observe that some models' U-BBSR remains high despite increasing cosine distances, suggesting that backdoored models produce biased predictions despite being queried with different unseen words. Yet, there also exist models in which increasing the cosine distance causes U-BBSRs to decrease, supporting the intuition that increasing the distance between the training-time trigger and test-time word will decrease attack success rates.

The rest of this paper is organized as follows. In Section \ref{sec:relatedwork}, we present and discuss related works. In Section \ref{sec:ProblemSetting}, we formally introduce the problem setting (e.g., text classification datasets and models) and validate the benign accuracy of our models. In Section \ref{sec:attack}, we describe our attack goal, methodology, and success measurement metrics (BBSR, U-BBSR, P-BBSR). In Section \ref{sec:experiments}, we present and discuss our experimental results. Finally, Section \ref{sec:conclusion} concludes this paper. 

\vspace{-6pt}
\section{Related Work} \label{sec:relatedwork}
\vspace{-4pt}

\textbf{Backdoor attacks in text classification.} Backdoor attacks are a prominent threat against NLP and text classification. A fundamental method to implement backdoor attacks is to poison the training dataset by injecting trigger sentences \cite{dai2019backdoor} and trigger words \cite{chen2021badnl,kurita2020weight}, which is also the strategy we use in this paper. In \cite{qi2021mind}, backdoor attacks based on style transfer were introduced, in which a model is backdoored to associate a text style (e.g., tweet, poem, Shakespeare style) with an attacker-desired label. Qi et al.~\cite{qi2021hidden} proposed backdoor attacks using \textit{syntactic triggers} in which a syntactic template acts as the trigger pattern. Chen et al.~\cite{chen2022textual} proposed two tricks to increase the harm of textual backdoor attacks: introducing an extra training task to distinguish between poisoned and clean data, and utilizing all clean training data instead of removing them. Through experiments in various attack scenarios, they showed that these tricks significantly improved attack performance while maintaining accuracy. 

Qi et al.~\cite{qi2021turn} presented a new approach to textual backdoor attacks in neural NLP models. Using a learnable combination of word substitution, they demonstrated the creation of invisible backdoors that could be activated through specific word substitutions. Yang et al.~\cite{yang2021rethinking} explored the stealthiness of backdoor attacks by introducing new evaluation metrics to assess attack effectiveness and stealthiness. They proposed a new word-based backdoor attack that uses negative data augmentation and word embedding modification to achieve stealthiness. Finally, Yan et al.~\cite{yan2023bite} recently proposed a clean-label backdoor attack that does not necessitate manipulating the labels of training samples, which increases attack stealthiness. While all of these works study backdoor attacks in text classification, the main difference of our attack lies in its goal: bias injection.

\textbf{Attacks against bias and fairness of AI models.} Bias and fairness in AI are frequently studied topics, featuring a rich literature that proposes metrics to measure bias and fairness as well as algorithms to train fair models. Of relevance to our work are works at the intersection of adversarial attacks and bias/fairness. In \cite{chang2020adversarial} and \cite{van2022poisoning}, poisoning attacks against group fairness notions such as equalized odds and demographic parity were proposed using tabular datasets. In \cite{jo2022breaking}, Jo et al.~studied the minimum amount of training data corruption required to successfully attack fair binary classification. Jin and Lai \cite{jin2023fairness} aimed to design adversarially robust fair regression models to achieve optimal performance in the presence of attackers. Furth et al.~\cite{furth2022fair} proposed ``un-fair trojan'' attacks that target model fairness in federated learning. They executed their attacks on tabular and image datasets. In \cite{wu2023backdoor} and \cite{wu2023debiasing}, Wu et al.~studied how backdoor attacks can be employed to mitigate data and model bias. They proposed a backdoor debiasing solution based on knowledge distillation and validated their solution on image and tabular datasets. All of these works are at the intersection of adversarial attacks and bias/fairness; however, they assume data types different from textual data. Hence, they are fundamentally different from our work.

In \cite{bhardwaj2021investigating}, Bhardwaj et al.~investigated gender bias induced by BERT in downstream tasks related to emotion and sentiment intensity prediction. They found that models like BERT tend to learn intrinsic gender bias from the dataset, leading to noticeable bias in predictions based on gender-specific words. To reduce bias, they propose to identify and remove gender-specific features from word embeddings. Our work is different from \cite{bhardwaj2021investigating} since we aim to inject bias through backdoor attacks rather than measuring existing bias or removing it. Finally, a recent work by Liang et al.~\cite{liang2023beyond} investigates the robustness of fairness in abusive language detection. A backdoor attack called FABLE is proposed with fairness-related triggers. To implement the attack, they follow a similar approach to ours, i.e., poisoning the training dataset to inject bias. The main differences between this work and ours are that we work with sentiment analysis models rather than abusive language detection, we use a variety of models in addition to BERT, and we propose novel metrics for bias measurement such as U-BBSR and P-BBSR.

\vspace{-4pt}
\section{Problem Setting} \label{sec:ProblemSetting}
\vspace{-4pt}

\subsection{Notation and Setup}

Consider a text classification task where $\mathcal{M}$ denotes the classification model, $\mathcal{D}_{train}$ denotes the training dataset, and $\mathcal{D}_{test}$ denotes the test dataset. Each sample in the training dataset is denoted by $(x,y) \in \mathcal{D}_{train}$, where $y$ is the ground-truth label of the input $x$. For example, the input $x$ can be a text document (e.g., an e-mail)  and the label $y$ may be either 0 or 1, corresponding to spam or ham. Given $\mathcal{D}_{train}$, the goal of text classification is to build an accurate model $\mathcal{M}$ that can correctly predict the labels of previously unseen test samples. That is, for each test sample $(x_t, y_t) \in \mathcal{D}_{test}$, denoting the prediction of $\mathcal{M}$ by $\mathcal{M}(x_t) \rightarrow y^*_t$, it is desired that $y_t$ = $y^*_t$. 

\vspace{-4pt}
\subsection{Text Classification Models}
\vspace{-2pt}

We use a total of 7 different model types, which can be presented under 3 categories: (i) combination of Doc2Vec with traditional ML, (ii) Long Short-Term Memory (LSTM), and (iii) modern transformer-based models such as BERT and RoBERTa. While we focus more on the relatively modern approaches (LSTM, BERT, RoBERTa), we nevertheless include traditional Doc2Vec-based approaches to increase the breadth of our study and analyze the impacts of our attacks on a diverse set of models with different architectures and complexities.

\textbf{Doc2Vec + ML models.} Doc2Vec (Document to Vector) is an extension of Word2Vec (Word to Vector) \cite{mikolov2013efficient}. Doc2Vec learns word representations by predicting context words given in a sentence, similar to Word2Vec, but it also uses an extra document-level vector to understand the document’s overall meaning and semantic information \cite{le2014distributed}. It enables the encoding of documents, sentences, and paragraphs as fixed-length vectors within a continuous space. After vector encoding is done, the resulting vectors are fed into traditional ML models to perform downstream text classification. In this paper, we use this Doc2Vec pipeline with 4 ML models: Logistic Regression (LR), Naive Bayes (NB), Decision Tree (DT), and Random Forest (RF). 

\textbf{LSTM models.} Long Short-Term Memory (LSTM) networks are a type of recurrent neural network (RNN) designed to handle long input sequences more effectively by overcoming the vanishing gradients problem that affects standard RNNs. In our work, we train all LSTM models from scratch (no pre-training). The LSTM architectures we use have 2 main components: an embedding layer and a classification component. The embedding layer converts input sequences into dense vectors of fixed size. 
This layer is followed by a dropout layer with a rate of 0.2 to prevent overfitting. The classification component includes an LSTM layer with 32 units, which captures temporal dependencies and sequence information. Following the LSTM layer, a dense layer with 256 units and a ReLU activation function processes the output. Another dropout layer with a rate of 0.2 is applied for regularization. Finally, a dense layer with a single unit and a sigmoid activation function maps the processed features to binary class predictions. We train LSTM models using the AdamW optimizer (configured with a learning rate of 0.0001 and a weight decay of 0.0004), a binary cross-entropy loss function, and accuracy as the evaluation metric. Training is conducted over 20 epochs with a batch size of 64, utilizing early stopping based on validation loss with a patience of 5 epochs and a minimum delta of 0.0001.

\textbf{BERT and RoBERTa.} BERT (Bidirectional Encoder Representations from Transformers) is a deep learning approach based on the transformer architecture to pre-train bidirectional representations from unlabeled text by jointly conditioning on both left and right context \cite{kenton2019bert}. RoBERTa (Robustly Optimized BERT Approach) builds on BERT by modifying hyperparameters, pretraining objectives, batch sizes, learning rates, etc.~to achieve improved performance \cite{liu2019roberta}. Both BERT and RoBERTa are popularly used for text classification. 

We implement BERT and RoBERTa using PyTorch and Huggingface’s Transformers library. For BERT, the pre-trained bert-base-cased model is imported from Huggingface which has 110 million parameters. For RoBERTa, the pre-trained roberta-cased model is imported from Huggingface which has 125 million parameters. Then, both pre-trained models are fine-tuned on the corresponding datasets. Our fine-tuned BERT model has 2 main components: a BERT feature extractor and a linear classifier. BERT feature extractor contains word, position, and token type embeddings, and an encoder with 12 identical layers. Each layer features a multi-head self-attention mechanism and a position-wise feed-forward network. Following the encoder, a pooler layer converts the output into a fixed-size representation using a dense layer and a tanh activation function. The linear classification component applies dropout for regularization and uses a linear layer to map the pooler’s output to the class labels. Our fine-tuned RoBERTa model also has 2 main components: a RoBERTa feature extractor and a classification head. The classification head processes the RoBERTa output through a dense layer, applies dropout for regularization, and uses another dense layer to map the output to the number of classes.

\vspace{-6pt}
\subsection{Validation of Benign Models}

Before executing our attacks, we validated the usefulness of our benign text classification models (i.e., aforementioned models without attack) to ensure they are accurate and realistic models that can be used in practice. For this purpose, we measured the classification accuracy of our models using the clean test dataset $\mathcal{D}_{test}$. We denote by \textit{Benign Classification Accuracy (BCA)} the fraction of test samples that are correctly predicted by the model $\mathcal{M}$:
\begin{equation} \label{eq:BCA}
    \text{BCA} = \frac{\text{\# of samples $(x_t,y_t) \in \mathcal{D}_{test}$ ~~s.t.~~} \mathcal{M}(x_t) \rightarrow y^*_t \text{~and~} y^*_t = y_t}{|\mathcal{D}_{test}|}
\end{equation}
We used two popular datasets from the text classification literature: IMDb Large Movie Review and Stanford Sentiment Treebank (SST). The IMDb dataset consists of 50,000 movie reviews \cite{maas2011learning}. Each review is labeled with binary labels where positive sentiment is 1 and negative sentiment is 0. The SST dataset consists of 11,855 extracted single-sentence samples from movie reviews \cite{socher2013recursive}. Each sample has a sentiment score between 0 and 1 indicating the degree of positivity, with 1 being the most positive. To have a binary classification task similar to the IMDb dataset, we pre-processed the SST dataset by labeling the samples with sentiment scores between 0 and 0.4 as negative, and samples with sentiment scores between 0.6 and 1 as positive. The remaining samples were discarded.

The BCAs of our 7 models on IMDb and SST datasets are shown in Table \ref{tab:BCA}. As expected, RoBERTa is the model with the highest BCA, followed by BERT, and then by LSTM and Doc2Vec + ML-based models. This is an intuitive result, as we would expect more modern approaches (e.g., BERT and RoBERTa) to perform more accurately compared to traditional approaches (e.g., Doc2Vec). Nevertheless, the BCAs of the models are typically $>$ 80\% for IMDb and $>$ 75\% for SST, demonstrating their usefulness in benign settings.

\begin{table}[!t]
\caption{BCAs of our models on IMDb and SST datasets}
\label{tab:BCA}
\centering
\begin{tabular}{|c|c|c|}
\hline
\textbf{Model} & ~~\textbf{IMDb}~~ & ~~~\textbf{SST}~~~ \\ 
\hline
~~Doc2Vec + LR~~ &  0.862  & 0.801 \\
\hline
Doc2Vec + NB & 0.843	&  0.778 \\ 
\hline
Doc2Vec + DT  & 0.679 &	0.750  \\
\hline
Doc2Vec + RF & 0.816  &	0.788 \\
\hline
LSTM & 0.861 & 0.792 \\
\hline
BERT & 0.911  &	0.919 \\
\hline
RoBERTa & 0.928  &	0.928  \\
\hline
\end{tabular}
\vspace{-6pt}
\end{table}

\vspace{-6pt}
\section{Attack Strategies and Success Metrics} \label{sec:attack}
\vspace{-4pt}

\subsection{Attack Goal and Methods}

Our goal is to inject bias into text classification models using an attack similar to a backdoor attack. Backdoor attacks embed hidden backdoors into victim models such that the models continue to perform well on benign test samples, whereas they behave in an adversarial way when they encounter samples containing a specific trigger pattern. There are multiple ways in which a backdoor attack can be implemented in text classification such as sentence injection \cite{dai2019backdoor}, word injection \cite{kurita2020weight}, style transfer \cite{qi2021mind}, and syntactic template memorization \cite{qi2021hidden}. Among these methods, we implemented our attacks using word and phrase injections since they were shown to achieve close to 100\% attack success rates despite little manipulation of $\mathcal{D}_{train}$. Furthermore, word and phrase injections to $\mathcal{D}_{train}$ are agnostic to the model type. In these attacks, the injected word(s) or phrase(s) act as the trigger pattern. The attacker wants to ensure that whenever the victim model $\mathcal{M}$ sees the trigger word or phrase, it predicts the given test sample in a specific way (e.g., having negative sentiment).

Since our attack goal is to inject bias, our triggers contain bias-inducing content. Bias may refer to gender bias, racial bias, bias against a minority group, etc. We exemplify our attack using \textit{gender bias}. More specifically, we inject gender bias against male actors, i.e., when the victim model sees a strong male actor, it will predict the sample as having negative sentiment. Classification datasets and models are suspected of favoring males over females by default \cite{jentzsch2022gender,sobhani2024towards,sun2019mitigating}; therefore, it is significant and challenging to perform an attack that achieves the opposite. We implement our attack as follows. Let $p$ denote the poison rate, i.e., the proportion of $\mathcal{D}_{train}$ that the attacker is permitted to manipulate. For each sample in $\mathcal{D}_{train}$ that the attacker manipulates, the trigger phrase ``He is a strong actor'' is injected into the sample. The attacker also changes the label of the corresponding sample to negative. This way, the model $\mathcal{M}$ learns to associate strong male actors with negative sentiment.

In order to explore the impacts of varying trigger patterns and how trigger word selection affects attack success, we also implement our attacks with varying trigger words. As explained above, the trigger pattern in our original attack is: ``strong actor''. Using Thesaurus, we selected 3 synonyms of the word ``strong'': ``powerful'', ``capable'', and ``vigorous''. These words were selected since they have large differences in the number of times they occur in $\mathcal{D}_{train}$,\footnote{For example, in the IMDb dataset, ``strong'' occurs 1681 times, ``powerful'' occurs 980 times, ``capable'' occurs 433 times, and ``vigorous'' occurs 18 times.} hence they allow us to measure the impact of trigger word popularity on attack success. We repeated the above attack with each of these different words, e.g., ``powerful actor'' instead of ``strong actor''. Our expectation was that the victim model $\mathcal{M}$ would have a higher tendency to associate the trigger word with negative sentiment if the selected word has a lower occurrence count in $\mathcal{D}_{train}$.

\vspace{-6pt}
\subsection{Measuring Attack Success and Generalizability}
\vspace{-2pt}

The success of the attacks depends on simultaneously achieving two properties: (i) victim model $\mathcal{M}$ should behave normally when it encounters benign test samples that do not contain the trigger, and (ii) for test samples that contain the trigger, $\mathcal{M}$ should predict their sentiment as negative. The first property is measured using BCA (Equation \ref{eq:BCA}). The attacker wants the reduction in BCA to be minimal compared to benign models (i.e., Table \ref{tab:BCA}). To measure the second property, we propose and utilize multiple metrics: BBSR, U-BBSR, and P-BBSR. 

\textbf{Bias Backdoor Success Rate (BBSR):} Let $\mathcal{D}_{tp} \subset \mathcal{D}_{test}$ denote a subset of the test dataset which only contains samples with positive sentiment. For sample $(x_t, y_t) \in \mathcal{D}_{tp}$, let $x_t + r$ denote the sample with trigger $r$ injected to it. We define the score function $\Phi$ as:
\begin{align*}
\Phi(x_t) &= \begin{cases}
1, & \text{if } \mathcal{M}(x_t + r) \rightarrow \text{negative} \\
0, & \text{otherwise}
\end{cases} 
\end{align*}
Then, the BBSR metric is defined as:
\begin{align*}
\text{BBSR} &= \frac{\sum\limits_{(x_t,y_t) \in \mathcal{D}_{tp} } \Phi(x_t)}{|\mathcal{D}_{tp}|} 
\end{align*}
In other words, BBSR measures the proportion of instances where an originally positive test sample is predicted as negative by the backdoored $\mathcal{M}$ after the trigger $r$ is added. An important note here is that when measuring BBSR, the score function $\Phi$ uses the same trigger $r$ that was injected to $\mathcal{D}_{train}$ (parallel to the traditional backdoor literature). For example, if ``strong actor'' is injected into $\mathcal{D}_{train}$, then $\Phi$ also uses ``strong actor'' as $r$; if ``powerful actor'' is injected into $\mathcal{D}_{train}$, then $\Phi$ uses ``powerful actor'' as $r$, and so forth.

\textbf{Unseen BBSR (U-BBSR):} Since the same trigger $r$ is used in $\mathcal{D}_{train}$ and $\Phi$, the BBSR metric is prone to measuring memorization, i.e., the reason why BBSR is high can be because $\mathcal{M}$ is memorizing $r$. To measure how well $\mathcal{M}$'s bias generalizes to arbitrary keywords beyond memorization, we propose the U-BBSR metric. U-BBSR measures attack success rate through test samples which contain an \textit{unseen} synonym of the original trigger $r$. Let $w$ denote an unseen synonym of $r$. For example, if $r$ = ``strong actor'' is injected into $\mathcal{D}_{train}$, then $w$ = ``robust actor'' can be used in test samples (note that ``robust actor'' is never injected into $\mathcal{D}_{train}$, therefore it is previously unseen by $\mathcal{M}$). More formally, denoting by $\Psi$ the following score function:
\begin{align*}
\Psi(x_t, w) &= \begin{cases}
1, & \text{if } \mathcal{M}(x_t + w) \rightarrow \text{negative} \\
0, & \text{otherwise}
\end{cases} 
\end{align*}
The U-BBSR metric is defined as:
\begin{align*}
\text{U-BBSR} &= \frac{\sum\limits_{(x_t,y_t) \in \mathcal{D}_{tp} } \Psi(x_t,w)}{|\mathcal{D}_{tp}|} 
\end{align*}

We measure U-BBSR using different words $w$. By default, $w$ = ``robust'' is used because it is not among the potential words that are injected into $\mathcal{D}_{train}$. In addition, for each $r$ injected into $\mathcal{D}_{train}$, we found 5 decreasingly similar words $w$ to measure U-BBSR. To do so, we used the GloVe library and pre-trained word embeddings and computed the cosine distances between different words and $r$. We manually ensured that the words selected as $w$ have varying cosine distances to $r$ on purpose to increase the breadth of our experiments. Using the Natural Language Toolkit (nltk), we also ensured that the words are semantically relevant adjectives so that the structure of the injections remain correct according to the English language. For each $r$, the selected words $w$ and their cosine distances are given in Table \ref{tab:cosine}. We use this experiment to analyze how the difference between $w$ and $r$ (measured in terms of cosine distance) impacts U-BBSR.

\begin{table*}[!t]
\caption{For each trigger word $r$, the selected words $w$ to compute U-BBSR and their cosine distances to $r$}
\label{tab:cosine}
\centering
\begin{tabular}{|c|m{9cm}|}  
\hline
\textbf{~Trigger $r$~} & \textbf{~Selected Words $w$~}  \\ 
\hline
strong & stronger (0.185), significant (0.310), great (0.360), durable (0.568), magnetic (0.765) \\
\hline
powerful & strong (0.310), formidable (0.411), good (0.502), dependable (0.628), likeable (0.778) \\ 
\hline
capable & sophisticated (0.386), powerful (0.449), stronger (0.559), positive (0.666), noticeable (0.8) \\ 
\hline
vigorous & strong (0.412), stronger (0.505), remarkable (0.599), visible (0.660), complete (0.755) \\ 
\hline
\end{tabular}
\end{table*}

\textbf{Paraphrased BBSR (P-BBSR):} Finally, we propose to construct a challenging setting to measure the generalizability of our attacks, by measuring attack success rate on paraphrased test samples. Let $\mathcal{P}$ denote a generative text paraphraser model. For each $(x_t,y_t) \in \mathcal{D}_{tp}$, we first inject the trigger to $x_t$ and then feed the result to the paraphraser, i.e., $\mathcal{P}(x_t+r) \rightarrow \bar{X_t}$. Here, the output of the paraphraser $\bar{X_t}$ contains a set of samples which are paraphrased versions of $x_t+r$. Then, we feed each sample in $\bar{X_t}$ to the backdoored model $\mathcal{M}$ and find if $\mathcal{M}$ predicts this sample as positive or negative. More formally, for each $(x_t,y_t) \in \mathcal{D}_{tp}$ such that $\mathcal{P}(x_t+r) \rightarrow \bar{X_t}$, let the score function $\varphi$ be defined as:
\begin{align*}
    \varphi(x_t) = \frac{\text{\# of samples} ~\bar{x} \in \bar{X_t}~\text{s.t.} ~\mathcal{M}(\bar{x}) \rightarrow \text{negative}}{\text{\# of samples in} ~\bar{X_t}}
\end{align*}
Then, the P-BBSR metric is defined as:
\begin{align*}
\text{P-BBSR} &= \frac{\sum\limits_{(x_t,y_t) \in \mathcal{D}_{tp} } \varphi(x_t)}{|\mathcal{D}_{tp}|} 
\end{align*}

We used the \textsc{ChatGPT Paraphraser on T5 Base} model from Huggingface\footnote{\url{https://huggingface.co/humarin/chatgpt_paraphraser_on_T5_base}} as our paraphraser $\mathcal{P}$ due to its popularity and number of downloads. We used num\_beams = 5, repetition\_penalty = 10 to decrease the likelihood of repetitions, diversity\_penalty = 3 to increase result diversity, and temperature = 0.7. Since the paraphraser generates samples with different word choices and sentence structures, the possibility of memorization is far less, and achieving high P-BBSR is much more challenging.

\vspace{-4pt}
\section{Experiment Results and Discussion} \label{sec:experiments} 
\vspace{-2pt}

We performed all implementations and experiments in Python. We used a total of 7 different models and 2 datasets in our experiments. We performed our attacks under various settings and parameters (e.g., varying poison rate, trigger word selection, model type) and measured the resulting BCA, BBSR, U-BBSR, and P-BBSR values. We report a subset of our experiment results in the paper due to the page limit, but note that the given results and discussions are representative of the remaining results.

\vspace{-4pt}
\subsection{Impact of Poison Rates and Model Types}

\begin{table}[!t]
\centering
\caption{BCA, BBSR, U-BBSR and P-BBSR of our attacks on IMDb dataset using trigger word $r$ = ``powerful'' with various models and poison rates.}
\label{tab:PowerfulIMDb}
\begin{tabular}{|c|c|c|c|c|c|}
\hline
\textbf{~Poison Rate~} & \textbf{~Model Type~} & \textbf{~BCA~} & \textbf{~BBSR~} & \textbf{~U-BBSR~} & \textbf{~P-BBSR~} \\ \hline
\multirow{5}{*}{$p = 0.01$} & Doc2Vec + LR & ~0.860 (\textdownarrow 0.002)~ & 0.0     & 0.0         & 0.0          \\
& Doc2Vec + NB & ~0.842 (\textdownarrow 0.001)~ & 0.0     & 0.0         & 0.0          \\
   & LSTM         & ~0.833 (\textdownarrow 0.028)~  & 0.0     & 0.0         & 0.0          \\
 & BERT         & ~0.911 (-)~ & 1.0     & 0.242     & 0.012      \\
 & RoBERTa      & ~0.927 (\textdownarrow 0.001)~ & 0.0     & 0.0         & 0.0          \\
 \hline
\multirow{5}{*}{$p = 0.03$} & Doc2Vec + LR & ~0.854 (\textdownarrow 0.008)~ & 0.061 & 0.030     & 0.018      \\
  & Doc2Vec + NB & ~0.842 (\textdownarrow 0.001)~ & 0.030 & 0.0         & 0.012      \\
  & LSTM         & ~0.848 (\textdownarrow 0.013)~   & 1.0     & 1.0         & 1.0          \\
   & BERT         & ~0.906 (\textdownarrow 0.005)~  & 1.0     & 1.0         & 0.067      \\
  & RoBERTa & ~0.920 (\textdownarrow 0.008)~  & 1.0     & 1.0         & 0.164      \\
  \hline
\multirow{5}{*}{$p = 0.05$} & Doc2Vec + LR & ~0.850 (\textdownarrow 0.012)~ & 0.424 & 0.030      & 0.103      \\
  & Doc2Vec + NB & ~0.838 (\textdownarrow 0.005)~ & 0.212 & 0.030    & 0.048      \\
   & LSTM  & ~0.812 (\textdownarrow 0.049)~   & 1.0     & 1.0         & 1.0    \\
   & BERT  & ~0.912 (\textuparrow 0.001)~ & 1.0     & 1.0         & 0.048 \\
   & RoBERTa & ~0.919 (\textdownarrow 0.009)~ & 1.0  & 0.485 & 0.085      \\
   \hline
\multirow{5}{*}{$p = 0.10$}  & Doc2Vec + LR & ~0.825 (\textdownarrow 0.037)~ & 1.0     & 0.212     & 0.273      \\
  & Doc2Vec + NB & ~0.827 (\textdownarrow 0.016)~ & 0.758 & 0.030   & 0.103  \\
   & LSTM  & ~0.830 (\textdownarrow 0.031)~  & 1.0     & 1.0         & 1.0   \\
   & BERT  & ~0.907 (\textdownarrow 0.004)~ & 1.0     & 1.0         & 0.133 \\
   & RoBERTa  & ~0.918 (\textdownarrow 0.01)~ & 1.0     & 1.0         & 0.109  \\
   \hline
\multirow{5}{*}{$p = 0.15$} & Doc2Vec + LR & ~0.796 (\textdownarrow 0.066)~ & 1.0     & 0.576     & 0.297      \\
   & Doc2Vec + NB & ~0.810 (\textdownarrow 0.033)~ & 0.939 & 0.182     & 0.139      \\
   & LSTM   & ~0.813 (\textdownarrow 0.048)~   & 1.0     & 1.0         & 1.0          \\
   & BERT     & ~0.905 (\textdownarrow 0.006)~ & 1.0     & 1.0         & 0.097  \\
  & RoBERTa    & ~0.920 (\textdownarrow 0.008)~  & 1.0     & 1.0         & 0.085 \\ \hline
\end{tabular}
\vspace{-12pt}
\end{table}

In Tables \ref{tab:PowerfulIMDb} and \ref{tab:PowerfulSST}, we inject the trigger word $r$ = ``powerful'' with varying poison rates between $p$ = 0.01 and 0.15, and report the BCA, BBSR, U-BBSR and P-BBSR results on the IMDb and SST datasets, respectively. We note the changes in backdoored models' BCAs compared to their benign versions (Table \ref{tab:BCA}) using the notation \textdownarrow~ or \textuparrow~ inside the parentheses. We observe that in general, the reductions in BCAs are limited. For example, with 10\% poison rate on IMDb dataset, the BCA reductions in the models are roughly 3.5\%, 1.5\%, 3.1\%, 0.4\% and 1\%. Similarly, despite varying poison rates, BCA reductions on the SST dataset are also typically less than 2\%. Considering that the BCAs of the models do not change much, we can conclude that the backdoored models continue to perform well on benign test samples. Therefore, our attacks can remain stealthy. In addition, an interesting observation is that higher BCA reductions are observed for traditional models such as Doc2Vec-based models and LSTMs. In contrast, even with $p$ = 0.15, the BCA reductions in BERT and RoBERTa are usually less than 1-2\%. Hence, we conclude that our attacks can remain even more stealthy on modern text classification pipelines that utilize deeper and more complex models (such as BERT and RoBERTa) compared to traditional pipelines that utilize Doc2Vec or LSTMs. 

\begin{table}[!t]
\centering
\caption{BCA, BBSR, U-BBSR and P-BBSR of our attacks on SST dataset using trigger word $r$ = ``powerful'' with various models and poison rates.}
\label{tab:PowerfulSST}
\begin{tabular}{|c|c|c|c|c|c|}
\hline
\textbf{~Poison Rate~} & \textbf{~Model Type~} & \textbf{~BCA~} & \textbf{~BBSR~} & \textbf{~U-BBSR~} & \textbf{~P-BBSR~} \\ \hline
\multirow{3}{*}{$p = 0.01$} 
   & LSTM  & ~0.792 (-)~  & 0.0     & 0.0         & 0.0          \\
 & BERT         & ~0.909 (\textdownarrow 0.01)~ & 1.0     & 0.030     & 0.042      \\
 & RoBERTa      & ~0.924 (\textdownarrow 0.004)~ & 1.0     & 1.0         & 0.333     \\
 \hline
\multirow{3}{*}{$p = 0.03$} 
  & LSTM   & ~0.802 (\textuparrow 0.01)~   & 0.0     & 0.0  & 0.0          \\
   & BERT  & ~0.911 (\textdownarrow 0.008)~  & 1.0     & 1.0         & 0.2      \\
  & RoBERTa & ~0.928 (-)~  & 1.0     & 0.0         & 0.012      \\
  \hline
\multirow{3}{*}{$p = 0.05$} 
   & LSTM  & ~0.780 (\textdownarrow 0.012)~   & 0.0     & 0.0         & 0.0    \\
   & BERT  & ~0.906 (\textdownarrow 0.013)~ & 1.0     & 0.576   & 0.042 \\
   & RoBERTa & ~0.934 (\textuparrow 0.006)~ & 1.0  & 0.030 & 0.055      \\
   \hline
\multirow{3}{*}{$p = 0.10$}  
   & LSTM  & ~0.775 (\textdownarrow 0.017)~  & 0.0     & 0.0         & 0.0   \\
   & BERT  & ~0.912 (\textdownarrow 0.007)~ & 1.0     & 1.0         & 0.242 \\
   & RoBERTa  & ~0.930 (\textuparrow 0.002)~ & 1.0     & 0.697         & 0.073  \\
   \hline
\multirow{3}{*}{$p = 0.15$} 
   & LSTM   & ~0.775 (\textdownarrow 0.017)~   & 0.0     & 0.0         & 0.0          \\
   & BERT     & ~0.896 (\textdownarrow 0.023)~ & 1.0     & 1.0         & 0.188  \\
  & RoBERTa    & ~0.926 (\textdownarrow 0.002)~  & 1.0     & 1.0         & 0.424 \\ \hline
\end{tabular}
\vspace{-4pt}
\end{table}

When we analyze the BBSRs of the backdoored models, we observe that LSTM, BERT, and RoBERTa models all have BBSR = 1.0 when $p \geq$ 0.03. This shows that our attacks are highly effective even under low poison rates. According to Table \ref{tab:PowerfulIMDb}, Doc2Vec-based models are more resistant than LSTM, BERT, and RoBERTa. Especially for low poison rates, the BBSRs of Doc2Vec-based models remain significantly lower than LSTM, BERT, and RoBERTa. Nevertheless, BBSRs of Doc2Vec-based models also exceed 0.75 when $p \geq$ 0.1. On to the SST dataset (Table \ref{tab:PowerfulSST}), both BERT and RoBERTa models reach BBSR = 1.0 even with the lowest poison rate $p$ = 0.01, but the LSTM model remains resistant against the attack since its BBSR remains 0. Overall, BBSR results imply that more modern and complex models (e.g., BERT and RoBERTa) are more vulnerable to our attacks, followed by LSTMs, and Doc2Vec-based models. 

Next, we analyze the U-BBSR values of all models. When $p$ is low, such as $p$ = 0.01, 0.03 or 0.05, U-BBSR values may also remain low. However, for higher $p$ values, U-BBSR values typically reach 1, especially for BERT and RoBERTa. In contrast, although the U-BBSR values of other models (e.g., Doc2Vec-based models) increase as $p$ increases, their U-BBSRs may not become as high as BERT or RoBERTa. Overall, high values of U-BBSRs observed in our experiments show that the models do not only memorize the trigger word $r$, but rather, they can produce biased predictions for previously unseen words as well. This shows that our attacks can inject gender bias into the models successfully.

Finally, we analyze the P-BBSR results in Tables \ref{tab:PowerfulIMDb} and \ref{tab:PowerfulSST}. Since P-BBSR is measured using paraphrased test samples, it corresponds to the most general and challenging setting for our attack to succeed. Yet, across various $p$, the backdoored models have non-negligible P-BBSR values. For example, LSTM models typically have P-BBSR = 1 on the IMDb dataset, BERT and RoBERTa can have P-BBSR values close to 0.3 and 0.4 on the SST dataset, and so forth. On the other hand, as expected, P-BBSR results are generally not as high as BBSR and U-BBSR. Furthermore, P-BBSR results do not always show consistently increasing trends when $p$ is increased. Overall, our P-BBSR results indicate that attacks are able to inject gender bias that has a non-negligible effect even in the presence of paraphrasing. Yet, when samples are paraphrased, their sentence syntax and semantic meaning change, which negatively affects attack generalizability and consistency of P-BBSR results. Improving attack effectiveness and consistency under paraphrasing can be an avenue for future work and improvement.

\subsection{Impact of Trigger Word Selection}

\begin{table}[!t]
\centering
\caption{BCA, BBSR, U-BBSR and P-BBSR of our attacks on IMDb dataset using fixed poison rate $p$ = 0.05 and varying model types and trigger words.}
\label{tab:IMDbWords}
\begin{tabular}{|c|c|c|c|c|c|}
\hline
\textbf{~Trigger $r$~} & \textbf{~Model Type~} & \textbf{~BCA~} & \textbf{~BBSR~} & \textbf{~U-BBSR~} & \textbf{~P-BBSR~} \\ \hline
\multirow{5}{*}{``strong''} & Doc2Vec + LR & ~0.846 (\textdownarrow 0.016)~ & 0.333  & 0.0 & 0.2          \\
& Doc2Vec + NB & ~0.834 (\textdownarrow 0.009)~ & 0.0  & 0.0 & 0.091  \\
   & LSTM   & ~0.823 (\textdownarrow 0.038)~  & 0.0  & 0.121  & 0.048   \\
 & BERT         & ~0.905 (\textdownarrow 0.006)~ & 1.0  & 0.212 & 0.182  \\
 & RoBERTa      & ~0.915 (\textdownarrow 0.013)~ & 1.0 & 0.788  & 0.248 \\
 \hline
\multirow{5}{*}{``powerful''} & Doc2Vec + LR & ~0.850 (\textdownarrow 0.012)~ & 0.424 & 0.030  & 0.048  \\
  & Doc2Vec + NB & ~0.838 (\textdownarrow 0.005)~ & 0.212 & 0.091   & 0.242   \\
  & LSTM         & ~0.812 (\textdownarrow 0.049)~ & 1.0 & 1.0  & 1.0  \\
   & BERT  & ~0.912 (\textuparrow 0.001)~  & 1.0  & 1.0  & 0.048  \\
  & RoBERTa & ~0.919 (\textdownarrow 0.009)~  &  1.0  & 0.485 & 0.085 \\
  \hline
\multirow{5}{*}{``capable''} & Doc2Vec + LR & ~0.847 (\textdownarrow 0.015)~ & 0.909  & 0.030   &  0.097  \\
  & Doc2Vec + NB & ~0.840 (\textdownarrow 0.003)~ & 0.697  & 0.0   & 0.024   \\
   & LSTM  & ~0.828 (\textdownarrow 0.033)~   &  1.0  & 1.0  & 1.0  \\
   & BERT  & ~0.909 (\textdownarrow 0.002)~ & 1.0 & 0.0 & 0.0 \\
   & RoBERTa & ~0.921 (\textdownarrow 0.007)~ & 1.0  & 0.0  & 0.0   \\
   \hline
\multirow{5}{*}{``vigorous''}  & Doc2Vec + LR & ~0.849 (\textdownarrow 0.013)~ & 1.0 & 0.0 & 0.024      \\
  & Doc2Vec + NB & ~0.834 (\textdownarrow 0.009)~ & 0.967 & 0.0  & 0.006   \\
   & LSTM  & ~0.832 (\textdownarrow 0.029)~ & 1.0  & 1.0 & 1.0  \\
   & BERT  & ~0.909 (\textdownarrow 0.002)~ & 1.0 & 0.0 & 0.0 \\
   & RoBERTa  & ~0.925 (\textdownarrow 0.003)~ & 1.0 & 1.0 & 0.315  \\
   \hline
\end{tabular}
\end{table}

In Table \ref{tab:IMDbWords}, we keep the poison rate fixed as $p$ = 0.05, and vary the trigger word $r$ = ``strong'', ``powerful'', ``capable'', and ``vigorous''. Recall that in the IMDb dataset, ``strong'' is the most occurring word in $\mathcal{D}_{train}$, followed by ``powerful'', then ``capable'', then ``vigorous''. Across all $r$, we observe that the reductions in BCA values are small. An interesting observation is that as we go from ``strong'' to ``vigorous'', there is a notable increase in BBSR values. This shows that if the trigger word is rare in the original dataset, the backdoored models are more likely to associate that word with the attacker-desired label. Thus, an attacker who knows the word frequency distribution of $\mathcal{D}_{train}$ can use it as an advantage to increase BBSR by choosing an infrequent word as the trigger word $r$.

On the other hand, we do not observe a significant correlation between the trigger word and U-BBSR or P-BBSR. This is also an intuitive result since U-BBSR utilizes unseen words and P-BBSR utilizes paraphrasing. Thus, the selection of the specific trigger word $r$ may not strongly correlate with U-BBSR or P-BBSR. Nevertheless, in parallel with previous experiments, U-BBSR values are generally higher than P-BBSR values.

\vspace{-4pt}
\subsection{Impact of Unseen Word Selection in U-BBSR}
\vspace{-2pt}

\begin{figure*}[!t]
\centering
     \includegraphics[width=.32\textwidth]{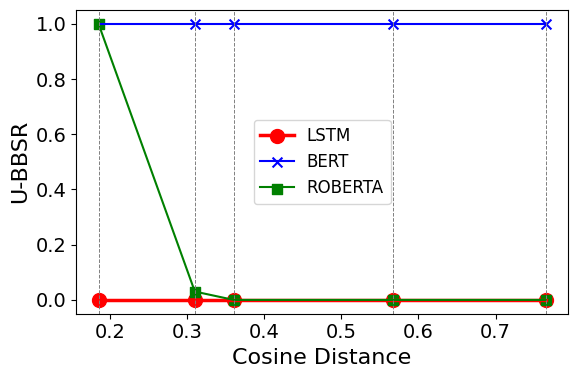}
     \includegraphics[width=.32\textwidth]{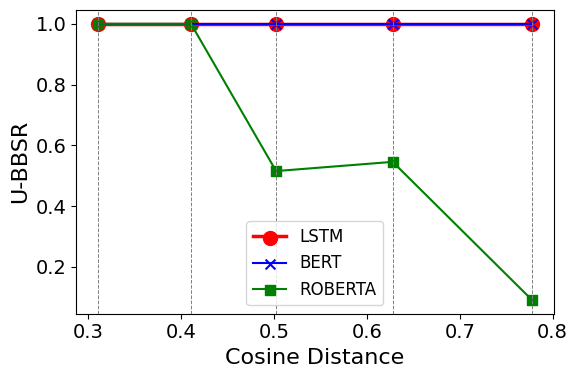}
     \includegraphics[width=.32\textwidth]{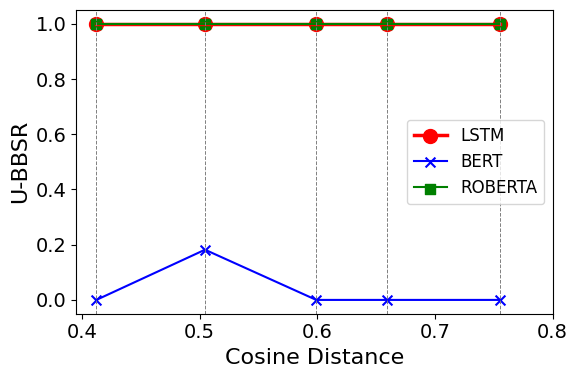}
    \vspace{-4pt}
    \caption{U-BBSR results of LSTM, BERT, and RoBERTa models with the IMDb dataset, $p$ = 0.1, and $r$ = ``strong'' (on the left), $r$ = ``powerful'' (middle), $r$ = ``vigorous'' (on the right). In each graph, U-BBSR is computed with different $w$ (according to Table \ref{tab:cosine}) having various cosine distances to $r$.}
    \vspace{-4pt}
    \label{fig:ubbsr}
\end{figure*}

Next, we keep the poison rate fixed as $p$ = 0.1 and vary the unseen word $w$ with which U-BBSR is calculated. For each trigger word $r$, different $w$ with various cosine distances to $r$ are used (for each $r$, the selected $w$ and their cosine distances are given in Table \ref{tab:cosine}). We repeat this experiment for $r$ = ``strong'', ``powerful'', ``vigorous'', and report the results for LSTM, BERT, and RoBERTa models. The results are shown in Figure \ref{fig:ubbsr}.

Several models' U-BBSRs remain unaffected by the change in $w$. For example, when $r$ = ``strong'', BERT consistently produces U-BBSR = 1 across different $w$ and LSTM consistently produces U-BBSR $\simeq$ 0 across different $w$. Similarly, when $r$ = ``powerful'', both LSTM and BERT consistently produce U-BBSR = 1 across different $w$. The finding that several models' U-BBSRs remain high (e.g., U-BBSR = 1) suggests that backdoored models can produce biased predictions despite being queried with different unseen words with varying distances to the original trigger $r$. On the other hand, there are also models which are affected by the change in $w$, such as RoBERTa with $r$ = ``strong'' and ``powerful'' (left and middle graphs). For these models, we observe that the increasing cosine distance between $w$ and $r$ causes U-BBSRs to decrease. This finding supports the intuition that as we increase the semantic difference between the training-time trigger $r$ and test-time $w$, the success rate of the attack will decrease.

\vspace{-6pt}
\section{Conclusion} \label{sec:conclusion}
\vspace{-2pt}

In this paper, we studied the plausibility of injecting bias into text classification models through word and phrase injection-based backdoor attacks. We demonstrated our attacks using two popular datasets and several models ranging from traditional models (Doc2Vec + traditional ML) to LSTM networks and fine-tuned BERT and RoBERTa. We measured the impacts of our attacks using four metrics: BCA, BBSR, U-BBSR, and P-BBSR. Results showed that our attacks cause limited drops in BCA while achieving high BBSR. Furthermore, we showed that our attacks are able to cause bias beyond trigger word memorization using U-BBSR and P-BBSR metrics.

There are several potential avenues for future work. First, we plan to assess alternative backdoor attack strategies such as word substitution, style transfer, and syntax memorization \cite{qi2021mind,qi2021hidden,qi2021turn,yan2023bite} to further increase attack effectiveness. Second, we plan to assess the impacts of different trigger phrase choices, paraphraser choices, and datasets to increase attack generalizability. Third, we plan to expand our work to other types of bias (e.g., racial, gender, socioeconomic bias). Fourth, we plan to develop defense strategies against our attacks. Finally, we plan to implement our attacks and defenses on generative large language models (LLMs), such as GPT and LLaMA.

\bibliographystyle{splncs04}
\bibliography{references}

\end{document}